\documentstyle[twocolumn,pre,aps,epsf]{revtex}

\def\d{{\rm d}}
\def\e{{\rm e}}
\def\vector#1{{\bf #1}}

\def\vk{{\vector k}}
\def\vq{{\vector q}}

\def\dps{\displaystyle}

\def\Tc{{T_{\rm c}}}

\def\kB{{k_{\rm B}}}

\def\TMTSFX{{${\rm (TMTSF)_2X}$}}
\def\TMTSFClO{{${\rm (TMTSF)_2ClO_4}$}}

\def\Tc{{T_{\rm c}}}

\def\lsim{\stackrel{{\textstyle<}}{\raisebox{-.75ex}{$\sim$}}}
\def\gsim{\stackrel{{\textstyle>}}{\raisebox{-.75ex}{$\sim$}}}

\def\kFxs{k_{{\rm F}x}^{(s)}}

\def\kFxsp{k_{{\rm F}x}^{(s')}}

\def\omegac{{\omega_{\rm c}}}

\begin{document}
\draft

\twocolumn[\hsize\textwidth\columnwidth\hsize\csname 
@twocolumnfalse\endcsname

\title{
Nodeless $d$-wave superconductivity 
in a quasi-one-dimensional \\
organic superconductor under anion order} 

\author{Hiroshi Shimahara} 

\def\runtitle{}

\address{
Department of Quantum Matter Science, ADSM, Hiroshima University, 
Higashi-Hiroshima 739-8526, Japan
}

\date{Received ~~~ December 1999}

\maketitle

\begin{abstract}
We propose a mechanism of nodeless $d$-wave superconductivity 
in a quasi-one-dimensional organic superconductor under anion order. 
It is shown that split of the Fermi surfaces due to the anion order 
eliminates the line nodes of a $d$-wave gap function 
on the Fermi surfaces, and a small finite energy gap appears 
in the quasi-particle excitations of superconductivity. 
We discuss that temperature dependences of thermal conductivity and 
NMR relaxation rate observed in a quasi-one-dimensional organic 
superconductor \TMTSFClO \hspace{0.5ex} can be explained consistently 
by the small energy gap. 
\end{abstract}


]

\narrowtext

Anisotropy of superconducting gap in organic superconductors 
is an important subject, since it is closely related to the pairing 
mechanism. 
In particular, a mechanism of an anisotropic superconductivity 
by pairing interactions mediated by antiferromagnetic (AF) 
fluctuations 
was proposed by Emery 
in the quasi-one-dimensional (Q1D) superconductors 
\TMTSFX~\cite{Eme86}, 
and has been studied by many authors, 
because of the proximity to the spin density wave. 
For example, it was shown that the phase diagrams and line nodes are 
semi-quantitatively reproduced by this mechanism in a microscopic 
calculation~\cite{Shi89}. 
At the present, however, there is not any experimental evidence that 
clarifies the origin of the pairing interactions in those compounds. 

For this subject, an NMR experiment was made 
in a Q1D superconductor \TMTSFClO \hspace{0.5ex} 
by Takigawa et al.~\cite{Tak87}. 
In their data, the nuclear relaxation rate $T_1^{-1}$ did not exhibit 
a coherence peak (Hebel-Slichter peak) just below the superconducting 
transition temperatue $\Tc$, and it was proportional to $T^3$ 
at lower temperatures. 
These results suggest that the superconducting gap function 
is anisotropic and has line nodes on the Fermi surface. 
By calculating the relaxation rate, Hasegawa et al. proposed 
two possibilities, 
singlet pairing with the gap function $\Delta(\vk) \sim \cos(k_y) + C$ 
(where $C \ll 1$) 
and triplet pairing with $\Delta(\vk) \sim \sin(k_y)$~\cite{Has87}. 
Here, we choose $x$-axis in the direction of the highly conductive chain, 
i.e., $a$-axis, and the $y$-axis in the direction of $b$-axis. 

However, recently, Belin et al. have observed a temperature dependence 
of thermal conductivity which indicates the nodeless gap~\cite{Bel97}. 
It may appear to contradict the NMR data at low temperatures, 
but as they pointed out, the NMR experiment was made only 
for $T \gsim 0.6 \, \Tc$, which may not be sufficiently low 
to determine the power of $T$. 
On the other hand, 
the absence of the peak of $T_1^{-1}$ just below $\Tc$ 
still suggests an anisotropic superconductivity. 
Hence, they argued the possibility of the triplet superconductivity 
with a gap function $\Delta(\vk) \sim \sin(k_x)$, 
which is anisotropic, but is finite over the whole open Fermi surfaces. 

For this triplet pairing, the Hebel-Slichter peak is much smaller 
than that for the $s$-wave pairing since the coherence factor 
is canceled in the momentum integral which gives the relaxation rate. 
However, in a theoretical prediction~\cite{Has87}, 
it does not seem to be small enough to explain the experimental data 
in which the peak is not seen at all. 
Further, although the NMR experiment was not made at low temperatures, 
a tendancy of $T_1^{-1} \propto T^3$ was seen above $0.6 \, \Tc$. 
Thus, if we believe the NMR data at the present, it is plausible 
that an anisotropic gap function which gives a quasi-particle density 
of states similar to that given by a gap function with line nodes 
developes at least for $T \gsim 0.6 \, \Tc$. 

In addition to the inexplicability in the NMR data, 
the following two points are not clarified in explanations based 
on the triplet pairing: 
(1) The triplet pairing seems to contradict that the proximity to the 
antiferromagnetism may favor the $d$-wave superconductivity; 
(2) If we assume the triplet pairing, a reentrant superconducting phase 
must be observed at a higher field above the upper critical field 
as Lebed' and Dupuis et al. calculated~\cite{Leb86,Dup93}, 
but there is not any experimental evidence of such reentrant phase 
at the present~\cite{Lee97}.

In this paper, we propose a mechanism of nodeless $d$-wave 
superconductivity which could explain the behaviors mentioned above, 
which appear to be inconsistent with each other. 
First, we assume pairing interactions mediated by the AF fluctuations. 
In the absence of anion order, the line nodes are obtained near 
$k_y = \pm \pi/2$, as in a result of a microscopic calculation~\cite{Shi89}. 
When anion order occurs, the lattice periodicity changes 
in the $b$-direction, 
and the Brillouin zone becomes half in the $k_y$-direction. 
Since the degeneracies are removed by potentials due to the anions, 
which are staggered in the $b$-direction, 
the Fermi surfaces split at the new zone boundary, 
near which the line nodes are situated. 
Hence, the line nodes are eliminated for appropriate parameters. 

Recently, Yoshino et al. estimated the staggered potentials 
due to the anion order $E_g \approx 0.083 \, t$ 
from their experimental data of magnetoresistance~\cite{Yos99}, 
where $t$ denotes the hopping energy along the chain. 
We will show below that the value of $E_g$ is, as an order of magnitude, 
large enough to eliminate the nodes of the $d$-wave superconducting gap 
near $k_y = \pm \pi/2$.

We start with a two-dimensional tight-binding Hamiltonian 
\def\eqH{(1)}
$$
     \begin{array}{rcl}
     H & = & \dps{ 
         - \sum_{i,j,\sigma} t_{ij} c_{i \sigma}^{\dagger} c_{j \sigma} 
         + \sum_{i,\sigma} \epsilon_i c_{i \sigma}^{\dagger} c_{i \sigma} 
         }\\
       &   & \dps{ 
         + \sum_{i,j} V_{ij} 
           c_{i \uparrow}^{\dagger} c_{i \uparrow} 
           c_{j \downarrow}^{\dagger} c_{j \downarrow} , 
         }\\
     \end{array}
     \eqno\eqH
     $$
with effective interactions $V_{ij}$. 
We define two sub-chains $A$ and $B$, and site energies 
$\epsilon_i = E_g$ and $-E_g$ for $i \in A$ and $i \in B$, respectively. 
The hopping energies $t_{ij}$ are assumed to be zero 
unless the sites $i$ and $j$ are nearest neighbor. 
For the nearest neighbor sites, 
$t_{ij} = t$ when $i$ and $j$ are on the same chain, 
while $t_{ij} = t'$ when $i$ and $j$ are on different sub-chains. 
We use units with $t = 1$.

\begin{figure}[htb]
\begin{center}
\leavevmode \epsfxsize=7cm  \epsfbox{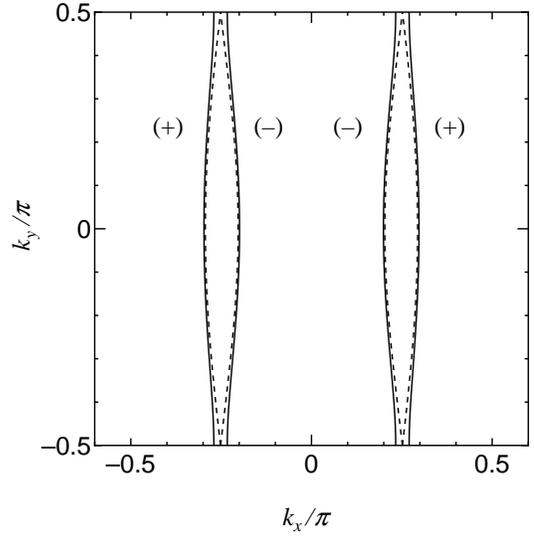}
\end{center}
\caption{Fermi surfaces of the $(+)$ and $(-)$ electron bands 
for 1/4 filling, when $t' = 0.1$. 
Solid and broken lines show those for 
$E_g = 0.083$ and $E_g = 0$, respectively.
} 
\label{fig:FS}
\end{figure}

The bilinear terms of the Hamiltonian are diagonalized by 
the transformations 
\def\eqcplusminus{(2)}
$$
     \alpha_{k,\sigma} =  
     \sum_{s = +,-} u_{\alpha s} c_{k,\sigma}^{(s)} 
     \eqno\eqcplusminus
     $$
with $\alpha = a, b$ and $s = +, -$, 
where we define $a_{i,\sigma} = c_{i,\sigma}$ for $i \in A$ and 
$b_{j,\sigma} = c_{j,\sigma}$ for $j \in B$. 
We obtain 
\def\equv{(3)}
$$
\renewcommand{\arraystretch}{2.0}
     \begin{array}{l} 
     u_{a +} = - u_{b -} = u(k_y) 
     = {\Bigl [} \frac{1}{2} {\Bigl (} 1 + \frac{E_g}{E_y} 
       {\Bigr )} {\Bigl ]}^{1/2} 
     \\
     u_{a -} = u_{b +} = v(k_y) 
     = {\Bigl [} \frac{1}{2} {\Bigl (} 1 - \frac{E_g}{E_y} 
       {\Bigr )} {\Bigl ]}^{1/2} , 
     \\
     \end{array}
     \eqno\equv
     $$
and the dispersion relations of the $(s)$ electron bands 
\def\eqepsilons{(4)}
$$
     \epsilon^{(s)}(\vk) = - 2 t \cos(k_x) - s E_y(k_y) , 
     \eqno\eqepsilons
     $$
where $E_y = \sqrt{{\epsilon_y}^2 + {E_g}^2} $ 
and $\epsilon_y = - 2 t' \cos(k_y)$. 
These forms of the dispersion relations were used by Yoshino et al. 
to explain their experimental results~\cite{Yos99}. 
The Fermi surfaces are modified by the staggered potential 
as shown in Fig.\ref{fig:FS}. 
In particular, the Fermi surfaces split at the boundary 
of the new Brillouin zone ($k_y = \pm \pi/2$).

We write the Fourier transformation of $V_{ij}$ as 
\def\eqV{(5)}
$$
     V(\vq) = V_0 (q_x) + V_1 (\vq) 
     \eqno\eqV
     $$
with intrachain and interchain interactions $V_0$ and $V_1$ respectively. 
Since the spin fluctuations are enhanced by the Fermi surface nesting 
with nesting vector $\vq_m \approx (\pm \pi, \pm \pi/2)$, 
which is equivalent to $\vq_m \approx (0, \pm \pi/2)$ 
in the presence of the anion order, 
the spin susceptibility $\chi(\vq,\omega)$ has peaks at $\vq = \vq_m$. 
The peak width is very small in the $p_x$-direction, 
while it is broad in the $p_y$-direction~\cite{Shi89}. 
Therefore, the effective interactions due to the exchange of the spin 
fluctuations are of long range in the $a$-direction, 
but of short range in the $b$-direction~\cite{Note1}. 
Thus, we take interactions up to those between nearest neighbor chains 
\def\eqVinterchain{(6)}
$$
     V_1(\vq) = - v_1(q_x) \cos(q_y) . 
     \eqno\eqVinterchain
     $$
The functions $V_0(q_x)$ and $v_1(q_x)$ have very sharp peaks 
at $q_x = \pm \pi/2$, reflecting the behavior of $\chi(\vq,\omega)$.

Here, we make a simplification that the peak widths of $V_0(q_x)$ 
and $v_1(q_x)$ are similar, which is qualitatively justified when 
$t'/t \ll 1$. 
Further, in order to express the sharp peak, 
we employ Lorentzian form 
\def\eqfdef{(7)}
$$
     f(q_x) = 
     \sum_{q_{mx}=\pm \pi/2} 
     \frac{1}{(q_x - q_{mx})^2 + {q_0}^2} , 
     \eqno\eqfdef
     $$
and put $V_0 (q_x) = g_0 f(q_x)$ 
and $v_1(q_x) = g_1 f(q_x)$.

Since the peaks are situated at the nesting vectors which connect 
the $(+)$ and $(-)$ electron bands, 
the peak width corresponds to the width of the area near the Fermi surfaces 
in which the pairing interactions are efficient. 
Therefore, we can assume that important contributions to the 
superconductivity mainly come from the pairing of electrons near 
the Fermi surfaces as in the starndard weak coupling theory. 
Hence, we can put 
$\langle {c_{-\vk \downarrow}^{(\pm)}}^{\dagger}
         {c_{\vk \uparrow}^{(\mp)}}^{\dagger} \rangle = 0$ 
when $E_g$ and $t'$ are not too small, 
since $\vk$ and $-\vk$ can not be near the Fermi surfaces of 
$(\mp)$ and $(\pm)$ electron bands, respectively, at the same time. 
This approximation is appropriate qualitatively when $E_g \gg \kB T$, 
and the experimental result $E_g = 0.083 \, t$ 
in \TMTSFClO \hspace{0.5ex} satisfies this condition.

Hence, we obtain a gap equation 
\def\eqgapeq{(8)}
$$
     \Delta^{(s)}(\vk) = - \frac{2}{N} \sum_{\vk' s'}
     V^{(ss')}(\vk,\vk') W^{(s')}(\vk') \Delta^{(s')}(\vk') , 
     \eqno\eqgapeq
     $$
where we define 
\def\eqWEskdef{(9)}
$$
\renewcommand{\arraystretch}{2.0}
     \begin{array}{rcl}
     W^{(s')}(\vk') 
     & = & \dps{
       \frac{1}{2E^{(s')}(\vk')}
       \tanh(\frac{E^{(s')}(\vk')}{2\kB T} ) 
       }\\ 
     E^{(s')}(\vk') 
     & = & {
     \sqrt{ (\epsilon^{(s')}(\vk'))^2 + (\Delta^{(s')}(\vk'))^2 } ~ . 
     }
     \end{array}
     \eqno\eqWEskdef
     $$
The effective pairing interactions $V^{(ss')}$ are obtained as 
\def\eqKssdef{(10)}
$$
\renewcommand{\arraystretch}{1.5}
     \begin{array}{rcl}
     \lefteqn{ V^{(++)}(\vk,\vk') = V^{(--)}(\vk,\vk') }\\
     & = & 
     \frac{1}{2} 
     {\Bigl [}
       V_0(\vk - \vk') {\Bigl (} 1 + \frac{{E_g}^2}{E_y E_y'} {\Bigr )} 
     + V_1(\vk - \vk') \frac{|\epsilon_y {\epsilon_y}'|}{E_y E_y'}
     {\Bigr ]}
     \\
     \lefteqn{ V^{(+-)}(\vk,\vk') = V^{(-+)}(\vk,\vk') }\\
     & = & 
     \frac{1}{2} 
     {\Bigl [}
       V_0(\vk - \vk') {\Bigl (} 1 - \frac{{E_g}^2}{E_y E_y'} {\Bigr )} 
     - V_1(\vk - \vk') \frac{|\epsilon_y {\epsilon_y}'|}{E_y E_y'}
     {\Bigr ]} , 
     \end{array}
     \eqno\eqKssdef
     $$
where $E_y' = E_y(k'_y)$ and ${\epsilon_y}' = \epsilon_y(k'_y)$.

Since the contributions from the electrons near the Fermi surfaces are 
dominant as we discussed above, we replace the effective pairing 
interactions with those on the Fermi surfaces. 
Further, for the sharp peak of the pairing interactions, 
we introduce an effective cutoff energy $\omegac$. 
Thus, we rewrite the right hand side of eq.{\eqgapeq} as 
\def\eqsumint{(11)}
$$
\renewcommand{\arraystretch}{1.5}
     \begin{array}{c}
     \dps{ 
     - \int_{-\pi/2}^{\pi/2} \frac{\d k_y}{\pi}
             \rho^{(s')} (0,k'_y) 
             \int_{-\omegac}^{\omegac} \!\!\! \d \epsilon \,\, 
             W^{(s')}(k'_x,k'_y)
     } \\
     \dps{ 
             \times 
             K^{(ss')}(k_y,k'_y) 
             \Delta^{(s')}(k'_y) , 
     } 
     \end{array}
     \eqno\eqsumint
     $$
where we define 
\def\eqKDkydef{(12)}
$$
\renewcommand{\arraystretch}{1.5}
     \begin{array}{rcl}
     K^{(ss')}(k_y,k'_y) 
     & \equiv & 
     V^{(ss')}((k_x,k_y),(\kFxsp(k'_y),k'_y)) \\
     && + V^{(ss')}((k_x,k_y),(-\kFxsp(k'_y),k'_y)) \\
     \Delta^{(s)}(k_y) 
     & \equiv & 
     \Delta^{(s)}(\pm \kFxs(k_y),k_y) . 
     \end{array}
     \eqno\eqKDkydef
     $$
Here, the function $\kFxs(k_y) > 0$ gives the value of $k_x$ on 
the Fermi surface of the $(s)$ electron band at $k_y$. 
The function $\rho^{(s)}(0,k_y)$ 
expresses the density of states at $\epsilon = 0$ and $k_y$ 
of the $(s)$ electron band of $k_x > 0$, 
which is calculated as 
$\rho^{(s)}(0,k_y) = 1/|4\pi t \sin(\kFxsp(k_y))|$.

Therefore, in the limit of $T \rightarrow \Tc$, 
we obtain the linearized gap equations in a form of eigen value equations 
\def\eqLGE{(13)}
$$
     \begin{array}{rcl}
     \lambda \Delta^{(s)}(k_y) 
     & = & \dps{ - \sum_{s'=\pm}
     \int_{-\pi/2}^{\pi/2} \frac{\d k'_y}{\pi}
     \rho(0,k'_y) } \\
     && \dps{ 
     \times 
     K^{(ss')}(k_y,k'_y) 
     \Delta^{(s')}(k'_y) , }\\
     \end{array}
     \eqno\eqLGE
     $$
where $\lambda = 1/\log(2 \e^{\gamma} \omegac/\pi \kB T)$ 
with the Euler constant $\gamma = 0.57721\cdots$. 
The eigen solution which belongs to the maximum positive eigen value 
gives the momentum dependence of the gap function near $T = \Tc$.

We solve eq.{\eqLGE} numerically 
for $E_g = 0$, $0.083$, and $0.2$, when $t'=0.1$. 
The chemical potential $\mu$ is adjusted so that the filling of 
holes $n = 1/4$. 
Since it was found in a microscopic calculation~\cite{Shi89} 
that the range of the interaction are over 30 sites along the chain, and 
that interchain interactions are much weaker than intrachain interactions, 
we use the parameters $q_0/\pi = 0.01$ and $g_0/g_1 = 5$ as an example. 
The qualitative result that we will show below is not very sensitive 
to the values of these parameters.

\begin{figure}[htb]
\begin{center}
\leavevmode \epsfxsize=7cm  \epsfbox{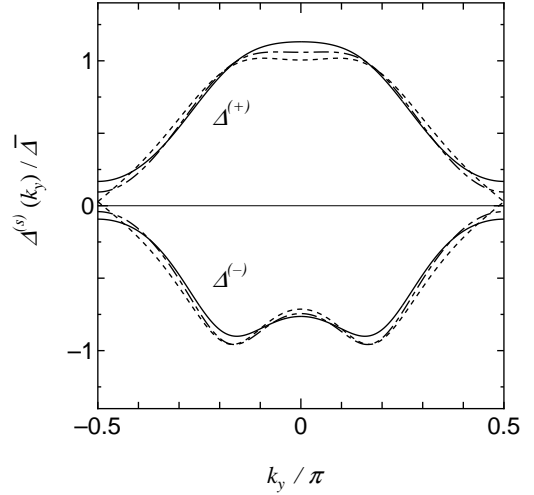}
\end{center}
\caption{The gap functions on the Fermi surfaces. 
The solid, dotted broken, and dotted lines show the solutions for 
$E_g = 0.2$, $0.083$, and $0$, respectively. 
}
\label{fig:Delta}
\end{figure}

Figure \ref{fig:Delta} shows the numerical result 
of $\Delta^{(s)}(k_y)$, where the normalization constant 
${\bar \Delta}$ is defined by 
\def\eqDeltabar{(14)}
$$
     {\bar \Delta} = 
     {\Bigl [}
     \int_{-\pi/2}^{\pi/2} \frac{\d k_y}{\pi}
     |\Delta(k_y)|^2 
     {\Bigr ]}^{1/2} , 
     \eqno\eqDeltabar
     $$
which must increase as temperature decreases. 
When $E_g = 0$, the result in the absence of anion order is reproduced. 
We have usual $d$-wave superconductivity with line nodes 
at $k_y \approx \pm \pi/2$ (dotted lines). 
The gap functions $\Delta^{(+)}$ and $\Delta^{(-)}$ have opposite signs 
equivalently to that $\Delta(|k_y| \lsim \pi/2) > 0$ and 
$\Delta(\pi/2 \lsim |k_y| < \pi) < 0$, 
which corresponds to the pairing with $\Delta(\vk) \sim \cos(k_y)$ 
in a more simplified model. 
When $E_g \ne 0$, we find that the line nodes vanish 
but the gap functions $\Delta^{(+)}$ and $\Delta^{(-)}$ still have 
opposite signs. 
The magnitudes $|\Delta^{(\pm)}|$ take minimum values at the zone boundary, 
and it is found that the minimum values increase with $E_g$.

\begin{figure}[htb]
\begin{center}
\leavevmode \epsfxsize=7cm  \epsfbox{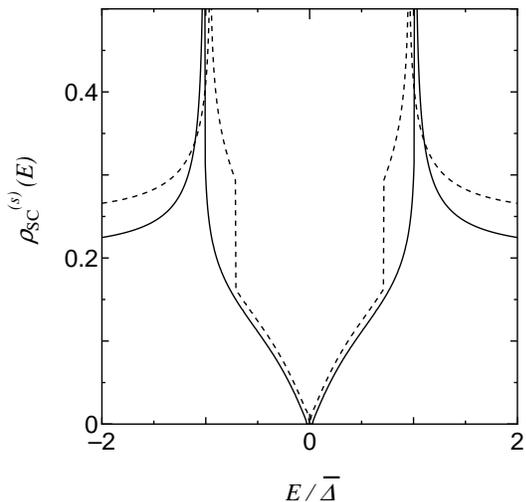}
\end{center}
\caption{Density of states near the Fermi level, when $E_g = 0$. 
Solid and broken lines show the results for the $(+)$ and $(-)$ 
electron bands, respectively. 
}
\label{fig:rho0}
\end{figure}

\begin{figure}[htb]
\begin{center}
\leavevmode \epsfxsize=7cm  \epsfbox{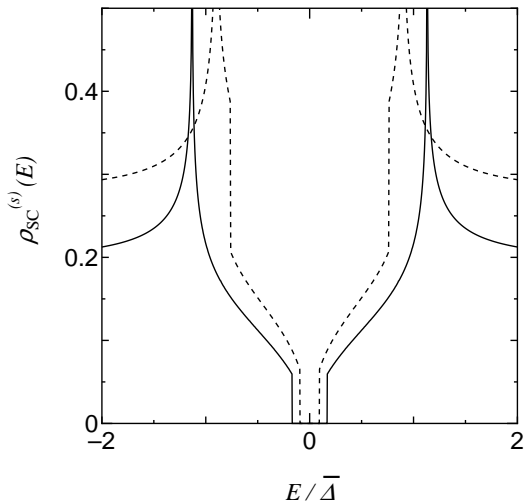}
\end{center}
\caption{Density of states near the Fermi level, when $E_g = 0.2$. 
Solid and broken lines show the results for the $(+)$ and $(-)$ 
electron bands, respectively. 
}
\label{fig:rho2}
\end{figure}

Figures \ref{fig:rho0} and \ref{fig:rho2} show the densities of states 
near the Fermi level for $E_g = 0$ and $0.2$, respectively. 
It is found that the energy gap opens when $E_g \ne 0$, 
but except the small energy gap near $E = 0$ 
the behavior of the density of states is very similar to that 
for $E_g = 0$.

The eigen value $\lambda$ can be regarded as an effective dimensionless 
coupling constant. 
The value of $\lambda t/g_0$ is estimated to be $29.9$, $25.1$, and 
$14.8$, for $E_g = 0$, $0.083 \, t$, and $0.2 \, t$, respectively. 
One could make an estimation of $\Tc$ if the values of $g_0/t$ and $\omegac$ 
are given, but in practice since an effect of self-energy renormalization, 
especially that from the pseudo gap, is very strong~\cite{Shi89}, 
we need a more careful treatment for a quantitative discussion.

In conclusion, we propose a mechanism of a nodeless $d$-wave 
superconductivity under the anion order. 
The momentum dependence of the gap function is calculated in a model 
with pairing interactions mediated by the AF fluctuations. 
It is found that the line nodes are located near the boundary of 
the half Brillouin zone in the absence of the anion order, 
and because of their locations, 
the line nodes vanish in the presence of the anion order.

Thus, the energy gap of the quasi-particle excitations 
of superconductivity becomes finite as shown in Fig.\ref{fig:rho2}. 
Such a small energy gap would not be observed very well in experiments 
at high temperatures near $\Tc$. 
Since the behavior of the density of states 
is very similar to that for $E_g = 0$ except the small energy gap, 
and the gap function changes its sign depending on $s = \pm$, 
temperature dependences near $\Tc$ of observed quantities are expected 
to be similar to those for the $d$-wave gap function with line nodes. 
In particular, it is expected that 
a Hebel-Slichter peak does not occur just below $\Tc$ 
and a tendancy of $T_1^{-1} \propto T^3$ is observed 
near and below $\Tc$, 
in NMR relaxation rate.

On the other hand, 
we could expect behaviors of a full gap superconductivity, 
for low temperatures. 
As temperature decreases, the superconducting gap developes, 
and finally the temperature becomes smaller than the finite energy gap. 
For such low temperatures, the electronic thermal conductivity 
should become vanishingly small.

These behaviors may explain the recent experimental results of the 
thermal conductivity and the NMR relaxation rate 
in \TMTSFClO \hspace{0.25ex} consistently. 
However, quantitative confirmation by fitting the experimental data 
seems to be difficult, because detailed informations 
of the pairing interactions, 
such as those in momentum, frequency, and temperature dependences, 
are needed for this purpose. 
In the future, observation of the Knight shift may clarify whether 
the pairing is singlet or triplet.

This work was supported by a grant for CREST from JST.


\end{document}